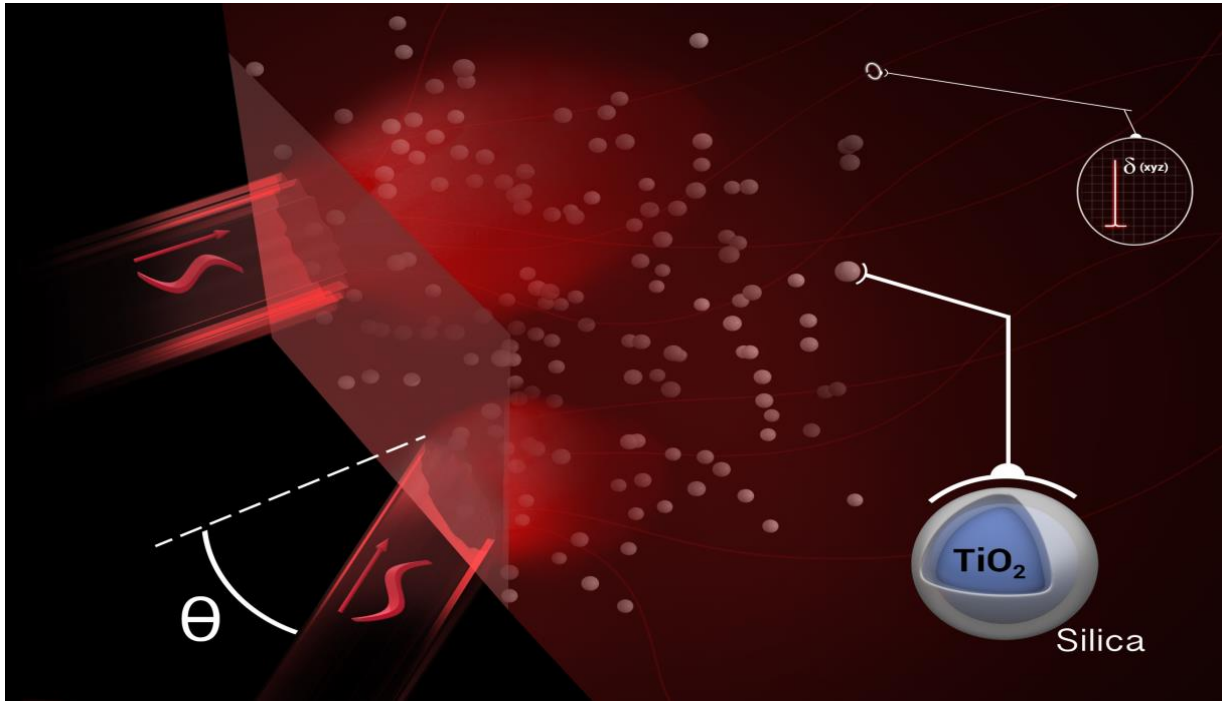

# Anomalous transport of light at the phase transition to localization: Strong dependence with incident angle

Ernesto Jimenez-Villar[1,2*], M.C. S. Xavier[2,3], Niklaus U. Wetter[1], Valdeci Mestre[4], W. S. Martins[5], G. F. Basso[6], V. A. Ermakov[7], F. C. Marques[7], G. F. de Sá[8]

[1] *Instituto de pesquisas Energéticas e Nucleares, CNEN_IPEN, São Paulo, SP 05508-000, Brazil*
[2] *Departamento de Física, Universidade Federal da Paraíba, João Pessoa, PB 58051-970, Brazil*
[3] *Departamento de Física, Universidade Estadual da Paraíba, Araruna, PB 58233-000, Brazil*
[4] *CCEA, Universidade Estadual da Paraíba, Patos, PB 58706-560, Brazil*
[5] *Departamento de Física, Universidade Federal Rural de Pernambuco, Recife PE 52171-900, Brazil*
[6] *Departamento de Informática, Universidade Federal da Paraíba, Joao Pessoa PB 58055-000, Brazil*
[7] *Departamento de Física Aplicada, Universidade Estadual de Campinas, Campinas SP 13083-859, Brazil*
[8] *Química Fundamental, Universidade Federal de Pernambuco, Recife, PE 50670-901, Brazil*
[*] *Corresponding author:* Ernesto.Jimenez@uv.es

Disordered optical media have seen a growing interest in recent year due to their potential applications in solar collectors, random lasers, light confinement and other advanced photonic functions. This paper studies the transport of light for different incidence angles in a strongly disordered optical medium composed by core–shell $TiO_2$@Silica nanoparticles suspended in ethanol solution. A decrease of optical conductance and an increase of absorption near the input border are reported when the incidence angle increases. The specular reflection, measured for the photons that enter the sample, is lower than the effective internal reflection undergone by the coherently backscattered photons in the exact opposite direction, indicating a non-reciprocal propagation of light. This study represents a novel approach in order to understand the complex physics involved at the phase transition to localization.

## 1. INTRODUCTION:

Anderson localization of light and associated phenomena have greatly attracted the attention of researchers in the past few decades [1–7]. Localization of light in a three–dimensional (3D) system (true Anderson localization) is an open research frontier in science that shows prospects of completely new optical phenomena, which might one day result in important photonics devices. However, direct observation of localization has shown to be difficult and elusive. Instead, we propose a strategy of observing the phase transition to localization by means of a set of specifically designed experiments. We demonstrate that the outcome of these experiments is completely different than expected in the diffusive regime and can be explained by the onset of localization. Localization (complete halt of transport) in 3D is extremely difficult to be obtained. The requirement for localization is known as the Ioffe–Regel criterion ($kl_T \sim 1$) [8], where $k=2\pi/\lambda$ and $l_T$ are the wave number and transport mean free path, respectively. However, a criterion for the phase transition to localization ($kl_T < x$ and, $x>1$) has not been clearly established. Notice that in realistic disordered optical media, composed of scatterers of size $\sim\lambda$, the interaction between scatterers (mean spacing < size) may lead to some degree of correlations in their positions [9]. The latter would imply that, in certain microscopic regions, $kl_T$ can reach lower values than the macroscopic $kl_T$ value (average) measured experimentally, being able to satisfy the strict criterion for localization ($kl_T \sim 1$). Consequently, both regions with localized and extended modes could coexist within a same sample (localization transition). Not–Brownian motions (out of equilibrium, sub–diffusive) of particles in colloidal suspensions have been previously predicted in spatially correlated random potentials [10,11].

Although this transition regime has been theoretically predicted in disordered electronic systems [12,13], its observation has proven elusive in optics, leading to a certain frustration of the optics community with respect to further investigations in this area [14]. Only very recently, it has been shown, through theoretical simulation, that a probable reason for this difficulty can be attributed to the type of scatterers used in these previous experiments, showing that a core–shell structure could be a promising strategy for reaching localization of light in 3D [15]. Because of the difficulty in observing directly localization of light, it is of paramount importance to discover new experiments that are a signature of localization of light. Based on the above arguments, we design for the first time a series of transport experiments as a function of the incidence angle to show the effects of the beginning of the critical regime of localization transition. Two scatterer concentrations, one operating in the purely diffusive regime [$14\times10^{10}$ NPs ml$^{-1}$] and one operating in the localization transition regime [$140\times10^{10}$ NPs ml$^{-1}$], were studied. Historically, various pioneering experiments that studied the transmission of electromagnetic waves through strongly disordered optical media have claimed the observation of localization of light [16–18]. However, these works were questioned firstly by opponents [19,20] and later refuted by their authors [21,22]. The inelastic scattering processes (absorption or nonlinearity) can lead to a decrease in the photon coherence length, hampering the interference effects (localization) [1,23]. In fact, according to the theoretical prediction of Sajeev John [1] and our previous experimental finding [6,24], an enhanced absorption arises when the system approaches localization. In a previous work [6], we reported several pieces of experimental evidence of localization transition in a strongly disordered optical medium composed by core–shell TiO$_2$@Silica nanoparticles (NPs) in ethanol solution. By using a Stöber method [25,26], TiO$_2$ NPs were coated with a homogeneous silica shell of ~40 nm thickness. The silica coating with thicknesses around or above 40 nm prevents the "optical" junction of the TiO$_2$ scattering surfaces (steric "optical" effect) [27], decreasing the near–field coupling that could hamper localization [28]. We called this property optical colloidal stability [27]. Additionally, the silica shell provides a light–coupling enhancement with TiO$_2$ scattering cores [29], inertness [30,31], and high dispersibility [32–35], which has enabled their use in numerous applications [36–38]. Transport experiments in this strongly disordered optical medium (TiO$_2$@Silica) showed an enhanced absorption when the system approached localization, from which an increase of the effective refractive index was proposed. This enhancement of absorption and refractive index was interpreted as that localized photons interact several times with the same particles, molecules or atoms within the localized state. The last phenomenon must be more pronounced near the input border, due to the increase of localization in the vicinity of the sample boundary for an internal reflection at the input border >0 as was theoretically predicted by Mirlin in disordered electronic media [39,40]. This can be understood as that the likelihood of a photon escaping from a hypothetical volume with dimensions of the order of the localization length should strongly depend on the mean reflectivity that photons would suffer at the borders that limit this volume. If this hypothetical volume is near the input border and the input surface of the sample forms part of this volume, the mean reflectivity that photons would suffer at these borders is always ≥0. However, for a volume that is completely within the scattering medium (away from the input surface), the mean reflectivity at the borders that limit this volume is equal to zero. The enhancement of the effective refractive index near the sample input border ($n_{eff0}$) by localization (successive elastic polarization of valence electrons to virtual states) finds a parallel in the dynamic barrier proposed by Campagnano and Nazarov [41] at the border of a disordered electronic medium. This means that the effective refractive index (internal reflection) that the localized photons would feel, would be higher than that felt by the photons that enter the sample (non–localized photons). Thereby, an increase of the incidence angle (internal reflection) should force the photons path to be longer (near the input border). Consequently, the likelihood of interference (near the input border) should increase. This issue was addressed theoretically by Ramos and co–workers [42], who demonstrated that the presence of a finite barrier (internal reflection) at the border provokes an increase of the quantum interference (localization increase) in a disordered electronic medium. Thereby, a decrease of the optical conductance would be expected when the incidence angle (internal reflection) increases. Clearly, this effect would only be appreciable if the system is at the localization transition such that an appreciable percentage of photons are localized, i.e. the density of localized states is comparable or higher than that of extended modes. Thereby, an increase of the density of localized states (localization increase) would provoke an appreciable decrease of the optical conductance. Furthermore, the effective refractive index near the input border would also be largely enhanced by localization itself, inducing an appreciable increase of the internal reflection with the incidence angle. For the diffusive regime (classical refractive index), there is a very low percentage of coherently backscattered photons (previously localized) and low contrast between the refractive indexes at the input interface (1.45–1.53 for silica–sample). Thereby, the internal reflection for the photons leaving the sample (coherently backscattered) would hardly change with the incidence angle. In the specific case that the internal reflection changes appreciably with the incidence angle, but the sample is at the purely diffusive regime (insignificant percentage of coherently backscattered photons, previously localized), an increase of the density of localized states with the internal reflection, would not appreciably affect optical conductance either, since the percentage of localized states in comparison with extended modes would be insignificant. In this paper, experiments of total and inelastic transmission, photon cloud propagation, average photon path length and absorption near the input border and coherent backscattering were performed, demonstrating strong influence of the incidence angle over optical conductance, absorption near the input border and the enhancement factor and width of the backscattering cone for the sample at higher [NPs]=[140x10$^{10}$ NPs ml$^{-1}$]. We demonstrate that the transport of light in the low concentration sample behaves insensitive with respect to the incidence angle (internal reflection), as expected at diffusive regime, while the high concentration sample shows a decrease of optical conductance and an increase of absorption near the input border. We remark that this anomalous behavior of transport of light near the mobility edge with the internal reflection has been theoretically predicted by Ramos and co–workers in disordered electronic system [42] but never shown in optics. These results could open new avenues for the design and manufacture of more efficient photonic devices based on strongly disordered optical media. For example, texturing of input surface, which increases the effective angle of incidence, would enhance notably the light–matter interaction (absorption) near the input border.

## 2. MATERIALS AND METHODS

**Sample preparation**
TiO$_2$@Silica NPs with a homogeneous silica shell of ~40 nm thickness, synthesized by an improved strobe method [25,26], were dispersed in ethanol solution at [140x10$^{10}$ NPs ml$^{-1}$]. For comparison, a sample with lower [NPs] in the diffusive regime [14 x10$^{10}$ NPs ml$^{-1}$] [6] was also prepared.

**Transmission experiment**
For all experiments, a CW He-Ne laser, model Uniphase *1125P* (10 mW, 633 nm), linearly polarized with polarization perpendicular to the incidence plane, was used. Total transmission is measured with an integrating sphere placed in contact with the back of the sample (fused silica cuvette). The laser spot size on the cell was <0.5 mm. The laser beam´s (He–Ne) incidence angles are 0º, 30º, 60º and 70º with regard to the normal of the cuvette, which correspond to incidence angles into the sample of 0º (0 mrad), 19.07º (333 mrad), 34.47º (600 mrad) and 37.89º (661 mrad), respectively. The reflection coefficients at the interface air–silica for incidence angles of 0º, 30º, 60º and 70º, are ~3.5%, ~5%, ~16% and ~28%, respectively. The specular reflection, measured at the interface silica–sample, was less than 1% for all incidence angles (negligible). For the experimental setup, see figure S1 of the supplementary material. The transmission coefficient (T(d)) is defined as the ratio between total transmitted flux and the incident flux and it was determined as a function of slab thickness (d) (Fig.1a).

**Propagation experiment**
The intensity profile *I*(x,y) of a probe beam (He–Ne laser) was measured for each incidence angle after propagating a distance d≈2.3 mm through the scattering medium. A CCD camera collected the image of the photon cloud at the sample output face. The diameter of the input probe beam is <100 μm full-width at half-maximum (FWHM). In order to obtain meaningful statistics, a total of 30 images, collected for different input points and intensities, were recorded for each incidence angle θ. For each incidence angle, the incident intensity entering the sample was corrected by the reflection coefficient at the air–silica interface (light entering the cuvette), which is ~3.5%, ~5%, ~16% and ~28% for 0º, 30º, 60º and 70º, respectively. For comparison, the propagation experiment was also performed for a sample with lower [NPs]=[14x10$^{10}$ NPs ml$^{-1}$] in the diffusive regime [6]. The experimental setup for the propagation experiment can be found in figure S2a of the supplementary material.

**Absorption experiments**
The macroscopic absorption length ($l_{MA}$) was determined from the exponential decay of the transmitted intensity I$_{TC}$ ∝exp(−d/$l_{MA}$) for large d using a very small solid detection angle (for experimental setup, see figure S3a in the supplementary material). In order to measure the average photon path length ($l_{eO}$) and absorption near the input border as a function of the incidence angle, the incident light, reflected by the samples, was measured with and without dye (Nile blue) [27,43,44]. The Nile blue (NIb) concentration is [1.5x10$^{-4}$M], which corresponds to a microscopic absorption length $l_{a(Nib)}$≈335 μm for 633 nm. We designated the ratio between the intensities reflected by the scattering medium with and without dye as the fraction of absorbed pumping (FAP). For this dye concentration, the macroscopic absorption length is $l_{MA(dye)}$≤10 μm. From the FAP measurements we can estimate the behavior of the average photon path length before being reflected or backscattered ($l_{eO}$) and the absorption near the input border (≤10 μm depth) when the incidence angle is increased. For the experimental setup see figure S3f in the supporting material.

**Backscattering experiment**
For the measurement of coherent backscattering, the sample is illuminated through a beam splitter that reflects 50% of the laser intensity and with a perpendicular polarization to the incidence plane. The light backscattered is collimated by a lens L$_3$ (25 mm focal length) and a CCD collects it. For the experimental setups, see figure S4 in the supplementary material.

## 3. RESULTS AND DISCUSSION

**Transmission experiment**
In order to study the transport of light as a function of the incidence angle, the transmission coefficient (T(d)) was determined for incidence angles (θ) of 0º, 30º, 60º and 70º with regard to normal incidence. Figure 1a shows T(d;θ) as a function of slab thickness *d* for θ equal to 0º, 30º, 60º and 70º. T(d;θ) can be fitted with a quadratic decay for all incidence angles T(d;θ)∝β(d$_0$+d)$^{-2}$, which would indicate localization transition [2,6,12]. d$_0$ is an experimental parameter introduced by Lagendijk and co–workers [45]. Owing to the light reflection at the interface air–silica (polarization perpendicular to incident plane), T(d;θ) tends to present different values for depth d=0 (T(0;θ)), showing lower T(0;θ) values for incidence angles of 60º and 70º. The T(d;θ) experimental points were corrected by scaling with the reflection coefficients at the interface air–silica measured for each incidence angle. After T(d;θ) correction for reflection at the interface air–silica, we extracted the derivative of (T(d;θ))$^{-1}$, ∂(T(d;θ))$^{-1}$/∂(d), from T(d;θ)= β(d$_0$+d)$^{-2}$ fittings for each incidence angle. For negligible absorption, the transport mean free path ($l_T$(d;θ)) is proportional to the inverse of the above derivative, ∂(T(d;θ))$^{-1}$/∂(d)∝[$l_T$(d;θ)]$^{-1}$. In figure 1b, the ratio: $[\partial(T(d;\theta))^{-1}/\partial(d)]/[\partial(T(d;0°))^{-1}/\partial(d)]$, which we denote as relative conductance with regard to the normal incidence $(G(d;\theta))$, is plotted as a function of slab thickness. Notice that for negligible absorption, $G(d;\theta)$ would represent effectively the inverse of the normalized conductance with regard to the normal incidence, since the conductance is proportional to the transport mean free path. For very large $d\to\infty$, $G(d\to\infty;\theta)$ tends to be an asymptotic value different for each incidence angle. This asymptotic value $(G(\infty;\theta))$ increases as the incidence angle is increased. Figure 1c shows the $G(\infty;\theta)$ increase when the incidence angle is increased. This fact might be explained through an increase of the internal reflection as the incidence angle is increased, which in turn, would lead to an increase of the density of localized states near the input border (superficial localized states) [42], i.e. an increase of localization near the input border. Notice that for a perpendicular polarization with respect to the incidence plane, the internal reflection increases continually as the incidence angle increases. Consequently, those photons with perpendicular polarization with regard to the incidence plane would be localized preferably (near the input surface) when the incidence angle is increased ($\theta > 0°$).

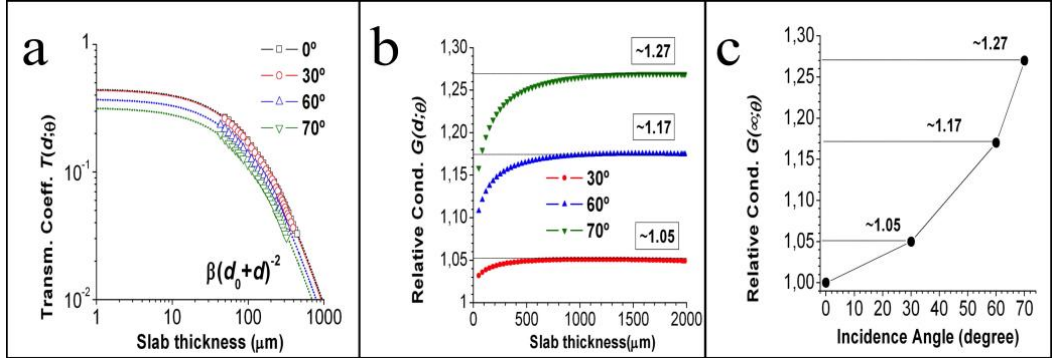

Figure 1. **For sample [140x10$^{10}$ NPs ml$^{-1}$], transmitted total intensity *vs* incidence angle.** a) Transmission coefficient for incidence angles θ of: 0°, 30°, 60° and 70° as a function of slab thickness (*d*). The black, red, blue and green dotted lines represent the fitting $\beta(d_0+d)^{-2}$ with experimental points for 0°, 30°, 60° and 70°, respectively. b) Relative conductance $G(d;\theta)$ as a function of *d*, c) Asymptotic values of relative conductance $G(\infty;\theta)$ as a function of the incidence angle.

From the above result, a decrease of optical conductance is inferred as the incidence angle is increased. Nevertheless, we must highlight that the data could be potentially affected by absorption, which would yield an inaccurate relative conductance, $G(\infty;\theta)$. Additionally, although we have carefully measured the residual stray light, the $T(d;\theta)$ values, extrapolated for large d by fitting, could be lightly spoiled by it. Therefore, in order to corroborate the above results (extrapolation for large d), an additional experiment of propagation was performed.

**Propagation experiment**
The intensity profile of a Gaussian probe beam was collected for each incidence angle after propagating a distance d≈2.3 mm through the sample. In order to corroborate the asymptotic values of the relative conductance $G(\infty;\theta)$, determined by extrapolation from the transmission experiment, the integrated intensity profiles $(I(\theta)=\int I(x,y)\partial x\partial y)$ were determined for each incidence angle θ. $I(\theta)$ values were corrected by the reflection at the input interface air–silica. In figure 2a, $I(0°)/I(\theta)$ ratios, which would represent the relative conductance $G(\infty;\theta)$ for a negligible absorption, are plotted as a function of the incidence angle. $G(\infty;\theta)$ shows a similar behavior to that extracted from the total transmission experiment. For incidence angles of 0°, 30°, 60° and 70°, normalized intensity profiles are displayed in figure 2d, 2e, 2f and 2g, respectively. The intensity profiles are not Gaussian; they could be fitted with a $\exp(-2(|r|/\sigma))^{1+\nu}$ function, where r is the radial distance of the beam center and $0<\nu<1$. Notice that the $\exp(-2(|r|/\sigma))^{1+\nu}$ function represents the overlap of the Gaussian and Poisson distributions, which is consistent with the localization transition regime (localized and extend modes coexisting). Notice that at localization, strong photon correlation at (x,y,d) emerges [40,46]. Therefore, photons from different points of the scattering medium must be strongly uncorrelated (Poisson law).
For the incidence angles of 60° and 70°, the intensity (profile) decreases more quickly for large r (red arrows pointing in figure 2f and 2g; 60° and 70°) and, for r near zero (cusp) the intensity profile adopts an acute form (discontinuous derivative), i.e. the intensity profile adopts a triangular shape. This effect could be caused by an increase of absorption near the input border when the incidence angle is increased, which in turn, would be originated by an increase of the density of superficial localized states. Notice that, the intensity for large r must represent those photons with longer paths, which would be those photons previously localized near the input border since localization must increase near the input border [39,40]. For each incidence angle, the confinement of the beam at the output plane is quantified by the inverse participation ratio:
$P \equiv [\int I(x,y)^2 \partial x\partial y]/[\int I(x,y)\partial x\partial y]^2 = 1/\pi [\int I(r)^2 \partial r]/[\int I(r)\partial r]^2$,
which has units of inverse area, and an effective width $\omega_{eff} = (P)^{-1/2}$. Figure 2b (left) shows the effective width as a function of the incidence angle, revealing an $\omega_{eff}$ decrease as the incidence angle is increased above 30°. Figure 2b (right) shows the relative effective width with regard to the normal incidence, $\omega_{eff}(0°)/\omega_{eff}(\theta)$ (normalized width). For 30°, $\omega_{eff}$ decreases less than 1%, which is within the measurement error. However, a significant $\omega_{eff}$ decrease is observed for 60° and 70°. This $\omega_{eff}$ decrease is associated to the quicker decay of the intensity profile at large r when the incidence angle is increased. For a negligible absorption, $\omega_{eff}$ is proportional to the conductance at large depth (~2.3 mm). Therefore, if $\omega_{eff}(0°)/\omega_{eff}(30°)\approx 1$, then for 30° and large depth→∞ (~2.3 mm), the conductance $\omega_{eff}(30°)$ should be equal to the conductance for 0° $\omega_{eff}(0°)$. Thus, let us introduce the following conjecture: For an incidence angle θ, the conductance for large depth, $\omega_{eff}(\theta)$, should correspond to the conductance at 0°, i.e. $\omega_{eff}(\theta)=\omega_{eff}(0°)$. This would imply that, for a negligible absorption (non–inelastic scattering), the conductance for large depth, $\omega_{eff}$, should not change as a function of the incidence angle, i.e. the density of localized states away from input border must be insensitive to the incidence angle. Moreover, a $\omega_{eff}$ decrease with the incidence angle would be directly related to an appreciable increase of the losses of light (absorption) near the input border, which would be caused by an increase of localization near the input border (increases of the light–matter interaction). This can be interpreted as that, for a negligible absorption, an increase of the density of superficial localized states (increases of localization near the input border), when the incidence angle is increased, would induce an increase of the total density of localized states throughout of the sample. However, the density of localized states for large depth (away from the input border) must remain unaltered. Notice that for large depths→∞, the influence of the input surface (internal reflection) becomes insignificant.

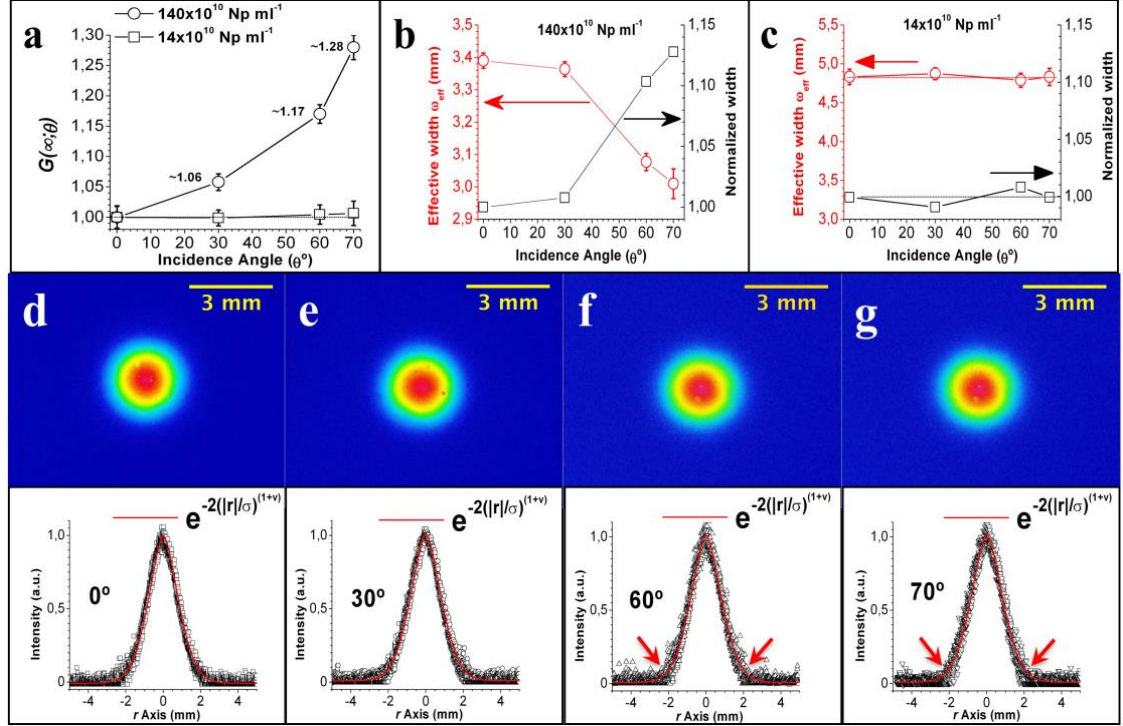

Figure 2. **Measurement of intensity profiles at the sample output face.** a) For [140x10$^{10}$NPs ml$^{-1}$] (localization) and [14x10$^{10}$NPs ml$^{-1}$] (diffusive regime), G(∞;θ)=I(0°)/I(θ) vs incidence angle, θ. b) For [140x10$^{10}$NPs ml$^{-1}$], (left–red) ω$_{eff}$ and (right–black) the relative effective width (normalized width) vs incidence angle. c) For [14x10$^{10}$NPs ml$^{-1}$] (diffusive), (left–red) ω$_{eff}$ and (right–black) the relative effective width (normalized width) are also plotted as a function of the incidence angle. The error bars are the statistic standard deviation of relative intensity and effective width (ω$_{eff}$). For [140x10$^{10}$ NPs ml$^{-1}$] (localization), normalized intensity profiles for incidence angles of: d) 0°, e) 30°, f) 60°, and g) 70°. Red arrows pointing quicker decay for large r. The intensity profiles are fitted to exp(−2(|r|/σ))$^{1+\nu}$ (red solid lines), where 0<ν<1.

For comparison, the propagation experiment was also performed for a sample in the diffusive regime with lower [NPs]=[14x10$^{10}$ NPs ml$^{-1}$] [6]. Figure 2a (open square) and 2c reveal, as expected, that both the integrated intensity (I(θ)) and effective width (ω$_{eff}$), respectively, are insensitive to the incidence angle.

**Absorption experiments**

From the latter results an increase of absorption near the input border was proposed. In order to estimate the influence of the incidence angle on absorption, the transmitted intensity (I$_{TC}$(d;θ)) was measured as a function of slab thickness for large d (between 100 μm and 400 μm) using a very small solid detection angle. The macroscopic absorption length (l$_{MA}$) can be determined from the inverse slope (log scale) of the exponential decay exp(−d/l$_{MA}$) [45]. An l$_{MA}$≈104 ±2 μm was found for all incidence angles, revealing that l$_{MA}$ for large d (away from the input border) is insensitive to the incidence angle. This latter can be interpreted as that the light–matter interaction away from input border remain unaltered when the incidence angle increases. In order to estimate the dependence of the conductance and absorption near the input border with the incidence angle, FAP measurements were performed as a function of the incidence angle. For the dye concentration used in this experiment [1.5x10$^{-4}$M], an effective macroscopic absorption length l$_{MA(dye)}$ ≤10 μm is estimated. Thereby, the absorption of reflected light should come from a layer (near the input surface) with thickness shallower than 10 μm (supplementary material). From the FAP values, we can estimate the average photon path length (l$_{eO}$) inside the scattering medium before being backscattered l$_{eO}$≈l$_{a(NIB)}$×ln(FAP) [27,43,44], which would yield us an estimative of the increase of light confinement near the input border (≤10 μm depth). An increase of the FAP value is observed as the incidence angle is increased (figure S3g supplementary material), reveling an increase of l$_{eO}$ and absorption near the input border as the incidence angle is increased. The latter can be understood as that, an increase in the incidence angle provokes an increase in the density of superficial localized states by the increase of the internal reflection [42]. In turn, an increase in the density of superficial localized states leads to an increase of the light–matter interaction [1,6], which would result in an increase of absorption ((l$_{In0}$)$^{-1}$) and refractive index (n$_{eff0}$) near the input border. Of course, l$_{In0}$ must be still longer than the microscopic coherence length, ξ$_{Coh}$ ≤ l$_{In0}$, for the absorption not to dominate the localization phenomenon. For comparison, FAP measurements were also performed for a sample in the diffusive regime with lower [NPs]=[14x10$^{10}$NPs ml$^{-1}$] [6], reveling that the FAP value is insensitive to the incidence angle (supplementary material, figure 3Sg).

In the above experiments, we show a decrease of transmitted intensity and infer an increase of localization and absorption near the input border as the incidence angle is increased. This fact was associated to an increase of the density of localized states near the input surface, which in turn was attributed to a large increase of the internal reflection (input border) felt by the coherently backscattered photons (previously localized), due to the enhancement of the effective refractive index near the input border. Therefore, the determination of this internal reflection, as a function of the incidence angle, becomes imperative in order to confirm our hypothesis.

**Backscattering experiment**

In order to determine experimentally the effective internal reflection, felt by the coherently backscattered photons ($IR$), the intensity of backscattering cone was measured as a function of the incidence angle. Figure 3a, 3b, 3c and 3d show the backscattering cone for incidence angles of 0º (0 mrad), 30º (524 mrad), 60º (1047 mrad) and 70º (1222 mrad), respectively. The specular reflection measured at the interface silica–sample for the photons that enter the sample is <1% for all incidence angles. From the intensity of the backscattering cone, we extracted the effective internal reflection felt by the coherently backscattered photons (previously localized) at the interface sample–silica (photons coming out the sample). Notice that the backscattering cone must represent those photons previously localized.

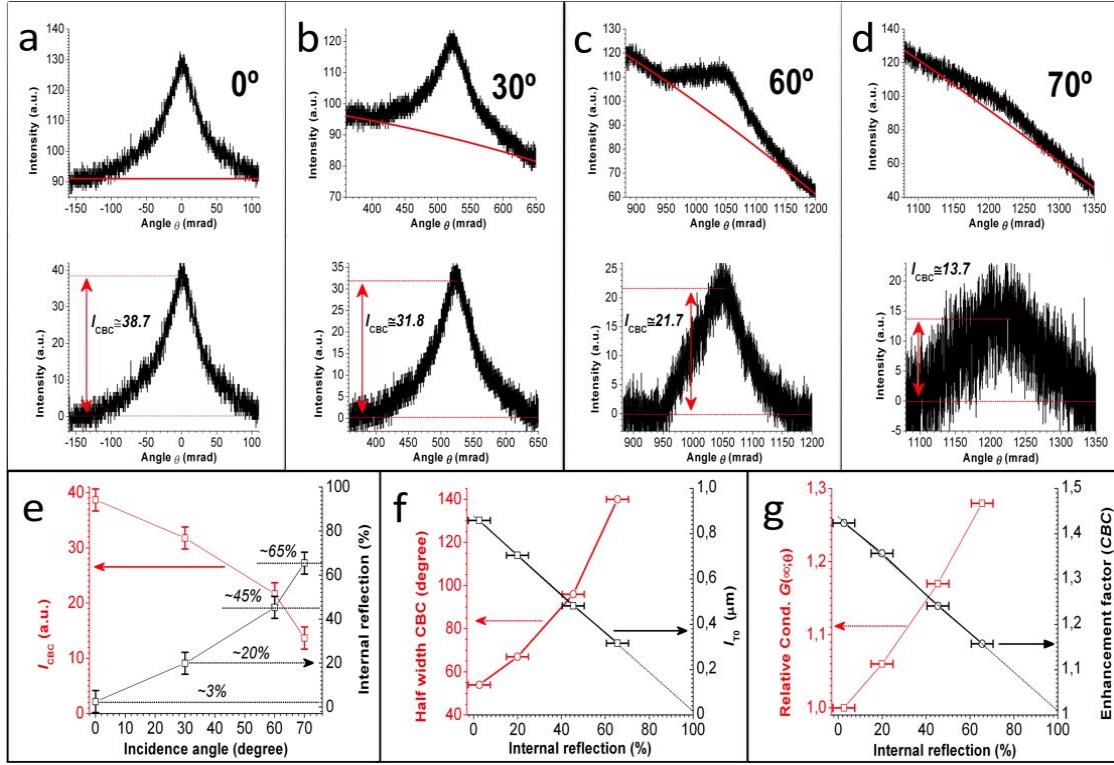

Figure 3. **For [140x10$^{10}$ NPs ml$^{-1}$] (localization regime), coherent backscattering cones for incidence angles of:** a) 0º, b) 30º, c) 60º and d) 70º. The red solid lines represent the background intensity (incoherently backscattered photons) taking into account the internal reflection at the interface silica–air (light coming out of the cuvette). The coherent backscattering cones obtained by subtraction of the background intensity are shown below each graph. e) (left–red) $I_{CBC}$ and (right–black) $IR$ (%) as a function of the incidence angle. f) (left–red) half width of backscattering cone and (right–black) $l_{T0}$ as a function of $IR$ (%). g) (left–red) asymptotic values of relative conductance (G(∞;θ)), extracted from the transmission and propagation experiments, and (right–black) enhancement factor of backscattering cone as a function of $IR$ (%). The black dotted lines at f) and g) represent linear fittings with the experimental points. Error bars correspond to the standard deviation of the intensity of the backscattering cone ($I_{CBC}$) and the calculated $IR$ (%).

The background intensity, represented by the red solid lines (figure 3a–d), was determined by calculating the internal reflection (Fresnel's equations) for the incoherently backscattered photons at the interface silica–air (supplementary material). The internal reflection at the interface sample–silica must be negligible for the incoherently backscattered photons (<1%), since the refractive indexes of sample and silica felt by these photons would be very close (1.53 and 1.45). The intensity of backscattered light was scaled by the reflection coefficients at the input interface air–silica (light entering the cuvette), which are ~3.5%, ~5%, ~16% and ~28% for 0º, 30º, 60º and 70º, respectively. The intensities of the backscattering cones were also rescaled by the internal reflection at the silica–air interface (photons coming out of the cuvette in the exact opposite direction). Figure 3e (left–red) shows the intensity of backscattering cone ($I_{CBC}$) and (right–black) $IR$ (%) as a function of the incidence angle. We calculated $IR$ for each incidence angle, $IR(\theta)$, by the expression $IR(\theta)=1-I_{CBC}(\theta)/I^*_{CBC}$, where $I_{CBC}(\theta)$ and $I^*_{CBC}$ are the intensity of backscattering cone measured for each incidence angle θ and the ideal intensity for null internal reflection, respectively. The internal reflection for the coherently backscattered photons at normal incidence ($IR(0º)≈3\%$) was determined considering an effective refractive index for depth≈0 of ~2 (supplementary material) [6]. From $IR(0º)≈3\%$, we can determine $I^*_{CBC}$ and, consequently, $IR(\theta)$ for the other incidence angles. $IR(\theta)$ values, determined for θ of 0º, 30º, 60º and 70º are ~3%, ~20%, ~45% and ~65%, respectively, which are considerably higher than the specular reflection measured for the photons that enter the sample (<1%) in the exact opposite direction. This indicates a non–reciprocal propagation of light, i.e. mirror–symmetry (parity symmetry) breaking. Notice that this large increase in the internal reflection undergone by the photons leaving the sample (coherently backscattered) would only be possible if the effective refractive index is largely enhanced. For a classical refractive index (1.53), the internal reflection for the photons leaving the sample (sample–silica interface) would be <1% for both polarizations and all incidence angles. We remark that absorption cannot cause such a decrease of the intensity of the backscattering cone, since the intensity decrease for the backscattered light with dye (absorption experiment near the input border, $l_{MA(dye)} \leq 10$ μm) for 0° and 70° is 10% and 32%, respectively. Therefore, without dye ($l_{MA} \approx 104$ μm$>10 \times l_{MA(dye)}$), the intensity losses (by absorption) of backscattered light should be considerably lower (<3%) for all incidence angles, which is within the measurement error. A few pioneering theoretical and experimental studies have addressed the mirror–symmetry breaking in photonic crystal cavities [47–49], however, no experimental evidence has been reported to date in a three–dimensional (3D) disordered optical medium. This phenomenon can be understood as that the photons that enter the sample feel a classical refractive index, but once they are localized; these photons feel an enhanced refractive index due to the successive elastic polarization of valence electrons to virtual states within the localized states. Figure 3f shows (left–red) the half width of backscattering cone and (right–black) the transport mean free path ($l_{T0}$), extracted by the half angle of backscattering cone [50], as a function of $IR$ (%). The width of backscattering cone increases monotonically as the incidence angle is increased, which is different to what is expected for a classical diffusive medium where $l_{T0}$ is insensitive to the incidence angle. $l_{T0}$ shows a decrease as $IR$ increases, tending to be zero for $IR \rightarrow 100\%$. A simple model for the internal reflection was taken into account for $l_{T0}$ correction (supplementary material) [51,52]. $l_{T0}$ values range from ~0.85 μm down to 0.3 μm, which represent $kl_T \sim 3–8 > 1$. Note that $kl_T \sim 1$ [8] is a general theoretical criterion for a complete halt of transport (complete localization). However, a clear criterion ($kl_T$) is lacking for the critical regime of localization transition. This is particularly true in a disordered optical medium composed of a colloidal suspension (TiO$_2$@Silica NPs), where the interaction between scatterers (repulsion) due to the electric field (ζ–potential=–75 mV) provided by the silica shell [53] can lead to certain correlations in scatterers positions. For a filling fraction of ~10.6% ([140x10$^{10}$ NPs ml$^{-1}$]), the mean separation between the scatterers is smaller than their size, which implies a strong interaction between the scatterers. Additionally, the repulsive force between the particles and its length range depends strongly on their size, whose polydispersity is 25%. In this way, an inhomogeneous distribution of scatterer positions should emerge at microscopic scale, leading to micrometric regions with $kl_T$ values that are lower and other regions with higher values than the averaged $kl_T$ value determined experimentally. Thereby, owing to the inhomogeneity at microscopic scale, localized and extended modes, coming from different regions with $kl_T$ values lower and higher than unity, respectively, can coexist in a same sample. This picture is what we have called in our previous works as the localization transition regime [6,7,24,43,54]. In this way, the average $kl_T$ value, extracted from the coherent backscattering experiment, would not provide a definitive criterion for the critical phase of localization transition. Furthermore, an abrupt phase transition from diffusive to localization regime, when scatteres concentration is increased (disorder increase), would be highly improbable, since for a realistic sample with scatterers highly concentrated and strong correlations in the scatterers positions $kl_T$ should cease to be homogeneous at microscopic scale.

Figure 3g (left–red) shows G(∞;θ) values, determined from transmission and propagation experiments, and (right–black) the enhancement factor of backscattering cone as a function of $IR$ (%). Notice that the enhancement factor tends to be 1 for $IR \rightarrow 100\%$, which was to be expected. These values of enhancement factor are considerably lower than expected for a linearly polarized probe beam (~1.8). This effect can be explained by: i) the effective refractive index felt by the coherently backscattered photons (previously localized) is considerably higher than that felt by the incoherently backscattered photons (non–localized photons), which leads to a higher internal reflection for the coherently backscattered photons; ii) the percentage of coherently backscattered photons with orthogonal polarization with regard to the original polarization could increase as the incidence angle is increased. We do not have a clear interpretation for this possible change of polarization. This could be explained by an anomalous nonlinear increase of refractive index, due to the intensity increasing (energy increase) within the localized states during the residence time of localized photons [43]. The latter would give rise to a phase shift that continuously increases during the photon residence time ($\tau_{eO}$). This phase accumulation can lead to interference breaking, emitting photons away from the localized state. This nonlinear increase of the refractive index in a localized state (closed loop path) would provoke an elliptic polarization, much like the Pockels effect. A similar nonlinear phenomenon was theoretically addressed by Buttiker and Moskalets (disordered electronic media) [55], who proposed that when the energy of the localized state changes, the localized state can emit non–equilibrium electrons and holes propagating away from the localized state within the edge state which acts similar to a waveguide. A detailed polarization study of the backscattering cone is called for, in order to determine the polarization of the coherently backscattered photons and its relationship with the incidence angle. The increase of localization with incidence angle near the input border could be also interpreted as that photons from superficial localized states that would be emitted by nonlinear effects (non–equilibrium) [43,55] can be again trapped in another superficial localized state, due to the increase of internal reflection. This latter implies in that the density and residence time (Q factor) of superficial localized states would increase as internal reflection (incidence angle) increases, which can be inferred from the theoretical predictions of Mirlin [39,40] and Ramos and co–workers [42] in disordered electronic media. Notice that an increase of the internal reflection with incidence angle would be remarkable, mainly for the coherently backscattered photons (previously localized), due to the enhanced refractive index that these photons would feel. For the incoherently backscattered photons, this effect would be considerably lower, since such photons would feel a classical refractive index. This means that the strong influence of the incidence angle over localization near the input border (conductance, absorption, refractive index), inferred from the above experiments, would only be appreciable if the percentage of localized photons is high (system is at localization transition), such that the conductance is strongly dependent on the density of localized states. Moreover, the effective refractive index would be largely enhanced near the input border, which in turn, would lead to an internal reflection strongly dependent on the incidence angle. In the diffusive regime, the conductance, absorption and refractive index near the input border are insensitive to the incidence angle. From the above results and ideas, we could infer that the dependence of the transmitted intensity (conductance) with the incidence angle can be described through the transport of light near the input border. The latter could be interpreted as the dependence of the light transport with the incidence angle being determined by the superficial localized states. Notice that, away from the input border, the mean reflection at the border of a hypothetical volume with dimensions around the localization length, ($\zeta_L$) would remain unchanged, equal to zero, as the incidence angle is increased. Thereby, the density of localized states (away from the input border) should remain unaltered. We remark that the above experimental results contradict the theoretical prediction of Skipetrov and van Tiggelen [56] that concluded that there is a decrease of localization near the border. Their conclusion is a result of the assumption that near the boundaries the waves could easily escape from the sample, reducing the probability of interference effects. However, we think that the important parameter for analyzing should not be the whole sample volume, but a volume with dimensions around the localization length, $\zeta_L$, where the photons are localized. Thereby, the probability to escape from a hypothetical volume ($\zeta_L$ dimensions) located completely within the sample would be higher than from a superficial hypothetical volume where the input surface forms part of this volume, since the internal reflection at the borders of a hypothetical volume completely within the sample is always equal to zero.

## 4. CONCLUSION

Core–shell $TiO_2$@Silica nanoparticles at [$140 \times 10^{10}$ NPs ml$^{-1}$] in an ethanol solution allowed us to study the strong influence of the incidence angle on the transport of light at localization transition. We remark that, owing to the scatterers interaction (mean spacing < size) and its size dependence, an inhomogeneous scatterers distribution at microscopic scale should emerge, which can sustain the critical regime of localization transition (Localized and extended modes coexisting). A decrease of conductance (localization increase) and an increase of absorption are reported near the input border as the incidence angle is increased. We remark that the values of relative conductance for 60º and 70º, extracted from above experiments, could be potentially affected by absorption, which
**6.**
## 7. APPENDIX: SUPPLEMENTARY MATERIAL

**Materials**
Ethanol alcohol HPLC with spectroscopic grade purity was supplied by MERCK, tetra-ethyl-ortho-silicate (TEOS) was supplied by Sigma–Aldrich, and the ammonia P.A. was supplied by MERCK. The titanium dioxide ($TiO_2$) with a rutile crystal structure was acquired from DuPont Inc. (R900). The $TiO_2$ grains have an average particle diameter of 410nm with a polydispersity of 25%. $TiO_2$ nanoparticles were coated with a silica shell of ~40 nm thickness via the Stöber method. In the first stage, 5 g of $TiO_2$ Nps were dispersed in 500 ml of absolute ethanol. This suspension was placed in an ultrasound bath for 20 minutes to disperse the particles and 6.67 mL of ammonia and 10 mL of TEOS were added. The TEOS and commercial ammonia ($NH_4OH$ 28%-30%) were added alternately in 100 portions of 100 μl and 220 μl, respectively. The synthesized $TiO_2$@Silica nanoparticle suspension was rota–evaporated, dried in an oven at 70 ºC for 2 h, and re–dispersed in ethanol at would yield inaccurate values. The measurement of the intensity of the backscattering cone allowed us to determine the effective internal reflection felt by the coherently backscattered photons. From the experimental results, we inferred that an increase of the internal reflection (incidence angle) must provoke an increase of the density and residence time (Q factor) of superficial localized states, which is reflected in an increase of localization and absorption near the input border. The specular reflection at the interface silica–sample, measured for the photons that enter the sample, is considerably lower than the effective internal reflection determined for the coherently backscattered photons in the exact opposite direction, which indicates a breaking of the mirror–symmetry (parity symmetry). This latter opens a way to manufacture an all–optical diode. The results shown in this work could present important technological applications. For example, the texturing of input surface (increase of the effective incidence angle), which must lead to an enhancement of localization and absorption near the input border, could open an avenue for the design and development of photonic devices based in strongly disordered optical media.

## 5. ACKNOWLEDGEMENTS

We gratefully acknowledge financial support from FACEPE, CNPq, FAPESP (grants 2017/05854-9, 2017/10765-5 and 2012/10127-5) and INCT/INES (465423/2014-0). Thanks to Professors Luis Poveda, Jorge Gabriel, for their thoughtful suggestions and discussions. We appreciate useful experimental support of Professors Mario Ugulino and Cid de Araujo. We thank Engineer Ricardo Acosta, for his corrections to the text. We extend additional thanks to designer Pedro Silva for the graphical image.

**Author Contribution**
E.J.V., M.C.S.X., N.U.W., V.M., W.S.M., G.F.B. & G.F.S performed the experiments, E.J.V., M.C.S.X., V.M., V.A.E., & F.C.M. synthesized and characterized nanoparticles and prepared samples, N.U.W. & E.J.V. analyzed the results and wrote the manuscript and E.J-V guided the research.

**Conflict of interest**
The authors declare no competing financial interests.

[$140 \times 10^{10}$ Nps ml$^{-1}$], equivalent to a filling fraction of 10.6 %. Another sample with lower NPs concentration [$14 \times 10^{10}$ Nps ml$^{-1}$] (diffusive regime) was also prepared. The $\zeta$–potential value of the core–shell $TiO_2$@Silica NPs dispersed in ethanol, calculated from the electrophoretic mobility using the Henry´s approximation, was −75 mV, which implies an appreciable and long–range repulsive forces between scatterers.

**Measurement of transmission coefficient**
In order to study the transmitted total intensity for incidence angles of 0º, 30º, 60º and 70º, the transmission coefficient was measured as a function of slab thickness for each incidence angle. The transmission coefficient is defined as the ratio between the total transmitted flux and the incident flux. The transmitted total intensity is measured with an integrating sphere placed in contact with the back of the cell. Figure S1a shows the schematic diagram of this experimental setup. A laser beam (He–Ne, 633nm) was passed through a positive lens $L_1$ (200 mm focal length), in order to obtain the focus with its waist near the pinhole PH (600μm diameter). Another lens, $L_2$ (50 mm focal length), was positioned 150 mm away from PH, in order to focus the beam on the cell, FF. The spot size on the input face of the sample is less than 0.5mm. The signal was collected through a multimode optical fiber (200 μm), coupled to a spectrometer HR4000 UV-VIS (Ocean Optics) with a 0.36 nm spectral resolution (FWHM). A study of the transmission coefficient for incidence angles θ of 0º, 30º, 60º and 70º was performed as a function of slab thickness $d$. The beam polarization is perpendicular to the incident plane. The reflection coefficients at the air–silica interface for incidence angles of 0º, 30º, 60º and 70º, are ~3.5%, ~5%, ~16% and ~28%, respectively. The specular reflection, measured at the silica–sample interface, was less than 1% for all incidence angles. The transmission coefficient (T(d;θ)) can be fitted with a $β(d_0+d)^{-2}$ function for all incidence angles and, for normal incidence, it tends to be ~0.45 at d=0. This effect can be explained through two factors: i) the internal reflection at the output air–silica interface (cuvette), ii) the light collection geometry of the integrating sphere. Notice that the scattering medium is contained in a fused silica cell. Thereby, an important part of transmitted light is reflected at the silica–air output interface. Additionally, the distance from the sample-silica interface to the entrance aperture of integrating sphere is 6–6.5 mm and, the diameter of entrance aperture of integrating sphere is ~18mm. Thereby, the collection angle of the integrating sphere for light coming from the interface silica–air is 55º–60º with regard to the normal. Therefore, the collected intensity (coming out of the sample–silica interface) must come from angles less than ~30º (with respect to the perpendicular to the cell surface), due to the light refraction at the output interfaces sample-silica and silica–air.

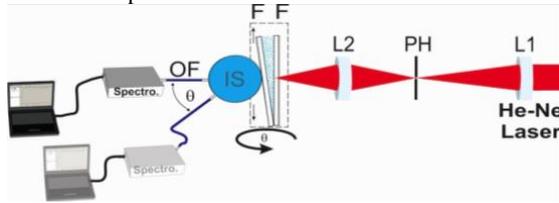

Figure S 1. Schematic diagram of the experimental setup for determination of transmission coefficient, L1 and L2, lens; PH, pinhole; F+F, cell consisting of two optical flat (fused silica) mounted on a translation stage; IS, integrating sphere is placed in contact with the back–cell; OF, optical fiber to collect the light in the spectrometer. A He-Ne laser beam with perpendicular polarization with regard to the incidence plane is introduced at different incidence angles, θ, with regard to the normal incidence (0º, 30º, 60º, 70º), which correspond to incidence angles into the sample of 0º (0 mrad), 19.07º (333 mrad), 34.47º (600 mrad) and 37.89 (661 mrad), respectively.

The total intensity (T(d)) that is collected can be expressed by the equation S1. $\vartheta$ is the collection angle with respect to the perpendicular to the cell surface, $\vartheta_1$ is the maximum collection angle and $f(\vartheta)$ is the angular dependence of the transmitted intensity. For an ideal case, $\vartheta_1$ is 90º (almost all scattered power is collected), however, in our case, $\vartheta_1$ is ~30º as stated above. From the equation S1, the quadratic decay of T(d) can be determined if it would be integrated over all angles (0º–90º).

$$T(d) = \beta^*(d_0+d)^{-2}\left[2\int_0^{\vartheta_1} f(\vartheta)d\vartheta\right] \qquad \text{S1}$$

$$\beta = \left[2\int_0^{\vartheta_1} f(\vartheta)d\vartheta\right]\beta^* \qquad \text{S2}$$

If β is defined as in S2, it implies that β values must be equal for both cases: ideal (0º–90º) and our collection (0º–30º). Notice that, for d>>$d_0$, $T_{91}(d)=T_{90°}(d)=β/d^2$. However, for ideal collections $d_0$ must approximately satisfy the relationship S3. Notice that, for d=0, $T_{91}(0)= β/(d_0^{91})^2$ and, for ideal case ($\vartheta_1$=90°), $T_{90°}(0)$ must tend to unity. $T_{30°}(0)$ can be expressed by the equation S4, where $T_{\parallel}(\vartheta)$ and $T_{\perp}(\vartheta)$ are the transmission coefficients at the silica–air interface for parallel and perpendicular polarizations, respectively. Thereby, $T_{30°}(0)≈40°/90°$, where 40° is the angle of total internal reflection at the silica–air interface.

$$d_0^{90°} = d_0^{30°} \times \sqrt{T_{30°}(0)} \qquad \text{S3}$$

$$T_{30°}(0) \qquad \text{S4}$$
$$= \int_0^{\sim 30°} \left(\frac{T_{\parallel}(\vartheta) + T_{\perp}(\vartheta)}{2}\right)$$
$$\times \cos\vartheta \, d\vartheta \approx 0.45 - 0.5 \approx \frac{40°}{90°}$$

In this way, for the configuration used in our experiment, T(d→0;θ) should tend to be around 0.45–0.5 for depth=0, which corresponds approximately to the value observed in the experiments. For incident angles of 60º and 70º, the transmission coefficient tends to be values appreciably lower than expected, which is the result of the losses of intensity by the reflection at the input interface air–silica (light entering the cuvette).

**Propagation experiment**
The intensity structure of a probe beam was studied after propagating a distance of ~2.3 mm through the samples. Figure S2 shows a schematic diagram of the experimental setup for this study. The linearly polarized probe beam (He–Ne laser) was passed through a positive lens $L_1$ (200 mm focal length) in order to obtain the focus with its waist near the pinhole PH (600μm diameter). Another lens, $L_2$ (38 mm focal length), was positioned 250 mm away from PH, in order to focus the beam on the cell, CV. The spot size on the input face of the sample is less than 100μm. Neutral density filters were used to attenuate the beam intensity (He–Ne). The cell consisted of two optical flats (fused silica, 3.2 mm thickness) separated by ~2.3 mm. In order to reduce the stray light, a metallic film with an aperture of ~5 mm diameter, through which the probe beam enters, is placed on the substrate at the silica–sample input interface. A CCD camera collected the images of probe beam at the output face. The probe beam was introduced at incidence angles of 0º, 30º, 60º and 70º. The beam polarization is perpendicular to the incident plane. In order to obtain meaningful statistics, a total of 30 images, collected for different input point and intensities, were recorded for each incidence angle. The integrated intensity profiles (I(θ)=∫I(x,y)∂x∂y) were determined for each incidence angle θ. For negligible absorption, the ratio of I(0°)/I(θ) must correspond to G(∞;θ) after correction of the losses of intensity by the reflection at the air–silica input interface. When d→∞>>$d_0$, the asymptotic value of the relative conductance G(∞;θ) is approximately equal to the I(0°)/I(θ) ratio, since

$$\frac{l_T(d;0°)}{l_T(d;\theta)} = \frac{\frac{\beta(0°)}{2(d+d_0(0°))}}{\frac{\beta(\theta)}{2(d+d_0(\theta))}} \approx \frac{\beta(0°)}{\beta(\theta)} \quad ; \quad \frac{I(0°)}{I(\theta)} \approx \frac{T(d\to\infty;0°)}{T(d\to\infty;\theta)} \approx \frac{\beta(0°)}{\beta(\theta)}.$$

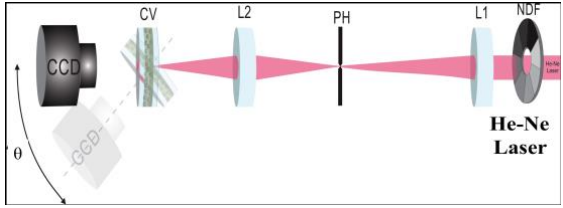

Figure S 2. Schematic diagram of the experimental setup for determination of the intensity profile after propagating through samples. L1 and L2, lens; PH, pinhole; CV, fused silica cuvette of ~2.3 mm optical pathlength; CCD camera; NDF, neutral density filter. At different angles of incidence, θ (0°, 30°, 60°, 70°), a He-Ne laser beam is introduced with perpendicular polarization with regard to the incidence plane.

For comparison, the propagation experiment was also performed for a sample in the diffusive regime with lower [NPs]=[14x10$^{10}$ NPs ml$^{-1}$]. As was expected, both the integrated intensity (I(θ)) and effective width (ω$_{eff}$) are insensitive to the incidence angle. This is because the transport of light in the diffusive regime must be insensitive of the angle of incidence.

**Absorption measurements**

Figure S3a shows a schematic diagram of the experimental setup for the measurement of the macroscopic absorption length ($l_{MA}$) at different angles of incidence. The laser beam (He–Ne) was passed through a positive lens L$_1$ (200 mm focal length) in order to obtain the focus near the sample surface (silica–sample input interface). The cell consisted of two optical flats (fused silica, 3.2 mm thickness), F, joined in a wedge and, therefore, the slab thickness depends on the height of the laser beam focus at the cell.

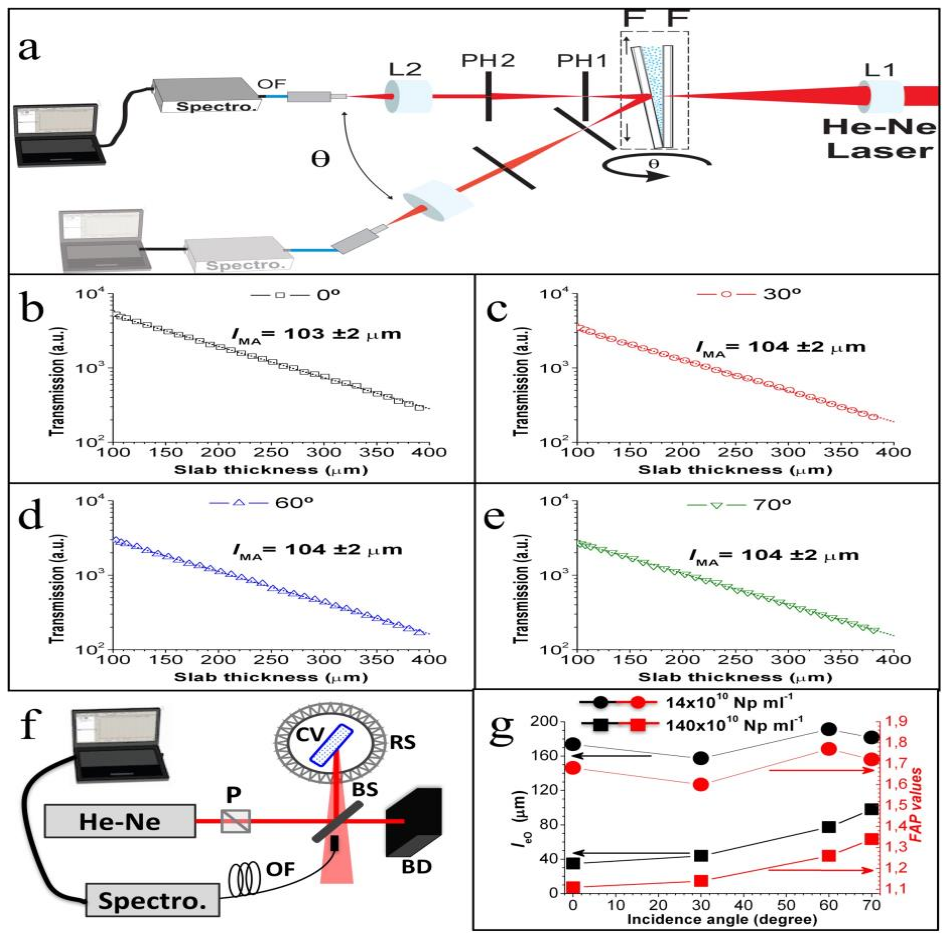

Figure S 3. a) Schematic diagram of the experimental setup for I$_{TC}$(d) determination as a function of slab thickness (d) for a very small detection solid angle. L1 and L2, lens; F+F, cell consisting of two optical flat mounted on a translation stage; PH1 and PH2, pinholes; OF, optical fiber to collect the light in the spectrometer. A He–Ne laser beam with perpendicular polarization with regard to the incidence plane is introduced at different incidence angles, θ, with regard to the normal incidence (0°, 30°, 60°, 70°). b–e) transmission curves for incidence angles of b) 0°, c) 30°, d) 60° and e) 70°. The black, red, blue and green lines represent the fitting with an exponential function exp(–d/$l_{MA}$) for the respective incidence angle. $l_{MA}$ values are displayed in each figure, showing to be insensitive to the incidence angle. f) Experimental setup for FAP measurement as a function of the incidence angle. The He–Ne laser is polarized perpendicular to the incident plane by a polarizer (P) and reflected by a beam splitter (BS) onto the sample (CV), which is mounted on a rotation stage (RS). The samples (CV), with and without dye, were rotated horizontally 30°, 60° and 70°; BD, beam dump; OF, optical fiber to collect the backscattered light in the spectrometer. g) Left–black and right–red represent $l_{eO}$ and FAP values, respectively, measured for [14x10$^{10}$ NPs ml$^{-1}$] (dots) and [140x10$^{10}$ NPs ml$^{-1}$] (squares), as a function of the incidence angle.

In order to measure the absorption, the transmitted intensity (I$_{TC}$(d)) for large d was measured as a function of slab thickness for four different angles of incidence (0°, 30°, 60° and 70°), using a very small solid detection angle. The laser spot size on the cell was <0.5 mm. In order to collect a very small solid angle, a pinhole PH$_1$ (600μm diameter) was positioned 20 mm away from the cell. Another pinhole, PH$_2$ (1200µm diameter), was also positioned 80 mm away from PH$_1$. Yet another lens, L$_2$ (50 mm focal length), allowed for focalization onto the optical fiber (OF). The multimode optical fiber (200 µm) was coupled to a spectrometer HR4000 UV-VIS (Ocean Optics) with a 0.36 nm spectral resolution (FWHM). Notice that, for a very small solid detection angle and large d, transmitted intensity should decay exponentially, $\exp(-d/l_{MA})$. In order to reduce the stray light, the fused silica plates (50 mm diameter) that form the wedge cuvette, were glued with opaque silicone glue. The stray light has been measured for all samples at large slab thickness (~1–2 order lower than signal), and it was subtracted from the I$_{TC}$ signal for each incidence angle. Figure S3 b–e show the exponential decay of the transmitted intensity at large d for incidence angles of 0º, 30º, 60º and 70º, respectively. From its inverse slopes, we obtain the macroscopic absorption length ($l_{MA}$). As can be observed (figure S3 b–e), $l_{MA}$ remains approximately constant ($l_{MA} \approx 104 \pm 2$ µm) as the incidence angle is increased, which means that the macroscopic absorption (for large d) is insensitive to the incidence angle. In order to study the dependence of conductance (localization) and absorption near the input border with the incidence angle, *FAP* measurements were performed as a function of the incidence angle by introducing [1.5x10$^{-4}$M] of dye (Nile blue) in the scattering medium. A He–Ne laser (633 nm) with perpendicular polarization with regard to the incidence plane is used. The samples (CV), with and without dye, were rotated horizontally 30º, 60º and 70º with respect to the normal incidence. The incident light, reflected by the samples, was measured with and without dye (Nile blue). We designated the ratio between the intensities reflected by the scattering medium with and without dye as the fraction of absorbed pumping (*FAP*). In this case, the dye acts as a "witness" that allowed us to estimate the average path length of the photons inside the sample before being backscattered. For this dye concentration, a microscopic absorption length $l_{a(Nib)}=335$ µm was measured, which must correspond approximately to a macroscopic absorption length of,

$l_{MA(dye)}=(l_{T0} \times l_{a(Nib)} \times 1/3)^{1/2} \approx (0.86µm \times 335µm \times 1/3)^{1/2} \leq 10$µm, for the incidence angle of 0º. In this way, from this *FAP* measurement, we can estimate the dependence with the incidence angle of $l_{eO}$ and absorption near the input border (≤10 µm depth). $l_{eO}$ is defined as the average photon path length inside the scattering medium before being backscattered. $l_{eO}$ can be expressed as $l_{eO} \approx l_{a(NIB)} \times \ln(FAP)$ [27,43,44]. For this calculus, we did not take into account the enhancement absorption factor ($\gamma_0$) by localization [6].

Figure S3f shows the experimental setup for the *FAP* measurement as a function of angle of incidence (0º, 30º, 60º and 70º). The samples (with and without dye) are illuminated through a beam splitter (BS). The light backscattered is collected by a multimode optical fiber (200 µm), OF, coupled to a spectrometer HR4000 UV–VIS (Ocean Optics) with a 0.36 nm spectral resolution (FWHM). Figure S3g shows, for [140x10$^{10}$ NPs ml$^{-1}$] (localization), the increase of $l_{eO}$ (left–black squares) and *FAP* (right–red squares) as the incidence angle is increased, which represents an increase of the average photon path length and absorption near the input border. For comparison, *FAP* measurements were also performed for a sample in the diffusive regime with lower [NPs]=[14x10$^{10}$NPs ml$^{-1}$]. Figure 3Sg shows that $l_{eO}$ (left–black dots) and *FAP* (right–red dots) values in the diffusive regime are insensitive to the incidence angle. We must highlight that at lower nanoparticles concentration the macroscopic absorption length is:

$l_{MA(dye)}=(l_{T0} \times l_{a(Nib)} \times 1/3)^{1/2} \approx (13.6µm \times 335µm \times 1/3)^{1/2} \approx 39$µm, which represents a depth of analyses higher than at [140x10$^{10}$ NPs ml$^{-1}$]. $l_{T0}=13.6$ µm was measured in our previous work [6].

**Measurement of the coherent backscattering cone**

The coherent backscattering cones were measured for incidence angles of 0º, 30º, 60º and 70º. The experimental setup used to this end is shown in figure S4. The linearly polarized laser beam (He–Ne) was passed through a positive lens L$_1$ (200 mm focal length) in order to obtain the focus with its waist near the pinhole PH (600µm diameter). Another lens, L$_2$ (150 mm focal length), was positioned 150 mm away from PH (focal length) in order to collimate the beam on the cell, CV. The sample is illuminated through a beam splitter that reflects 50% of the laser intensity. The backscattered light is collimated by a lens L$_3$ (25 mm focal length) and a CCD collects it. Neutral density filters were used to attenuate the beam intensity (He–Ne). The cell is composed of two fused silica optical flats (6 mm thickness). In order to average out the speckle pattern, the collection time was 500 seconds, which is enough for particle diffusion in the suspension. To collect the coherent backscattering cone at different angles of incidence, the sample was rotated horizontally 30º, 60º and 70º, which are equivalent to 523mrad, 1047mrad and 1222mrad, respectively. For an incidence angle of 0º, the sample was slightly tilted (horizontal) to keep the specular reflection from reaching the detector. The beam polarization is perpendicular to the incidence plane, which is the same polarization used in the transport and absorption measurements. The reflection coefficients for 0º, 30º, 60º and 70º are ~3.5%, ~5%, ~16% and ~28%, respectively. Owing to light refraction at the interfaces air–silica and silica–sample, the incidence angles into the sample are 0º (0 mrad), 19.07º (333 mrad), 34.47º (600 mrad) and 37.89 (661 mrad). The red solid lines in the figures 3a, b, c and d represent the background intensity of backscattering, which was determined taking into account the internal reflection at the silica–air interface (Fresnel's equations). Owing to the low contrast of refractive index, felt by the incoherently backscattered photons, at the sample–silica interface (1.53–1.45), the internal reflection at this interface was neglected. The multiple–backscattering background must be totally depolarized. Therefore, the background intensity was fitted through the equation S4.

$$\left(\frac{RC_\parallel(\vartheta) + RC_\perp(\vartheta)}{2}\right) \times \frac{\cos\vartheta}{\cos(\theta)} \times I_0 \quad\quad S4$$

Where $RC_\parallel(\vartheta)$, $RC_\perp(\vartheta)$, $\vartheta$, $(\theta)$ and $I_0$ are the reflection coefficients at the interface silica–air for the parallel and perpendicular polarization, the horizontal collection angle, the incidence angle (0º, 30º, 60º, 70º) and a constant that is associated to the incidence intensity (determined by the incidence angle), respectively. The intensity of backscattered light was scaled by the reflection coefficients for each incidence angle at the air–silica input interface (light entering the cuvette), which are ~3.5%, ~5%, ~16% and ~28% for 0º, 30º, 60º and 70º respectively. The intensity of backscattering cones was also rescaled by the internal reflection at the silica–air interface (photons coming out the cuvette in the exact opposite direction). To this end, we considered the classical refractive index of silica (1.45) and, clearly, a perpendicular polarization to the incidence plane.

The effective internal reflection (for the coherently backscattered photons) for the incidence angle of 0º was calculated through the Fresnel's equations considering that the coherently backscattered photons feel an effective refractive index of 2 near the input border (~3% internal reflection) [6]. Notice that, the proposed increase in the effective refractive index is connected with the known Kramers–Kronig relations, since the enhanced absorption coefficient, $\alpha_{FF0}(\omega)$, can be expressed as: $\alpha_{FF0}(\omega)\approx\gamma_0\times\alpha_0(\omega)$. Owing to the transport of light, measured in our previous work [6], which is approximately similar for a broad range of frequencies (532 nm to 633 nm), we consider an enhanced absorption factor approximately constant in $\omega$. Thereby, $n_{eff0}(\omega_0)=1+\gamma_0(n_{eff}(\omega_0)-1)$, where $n_{eff}(\omega_0)$ is the classical refractive index.

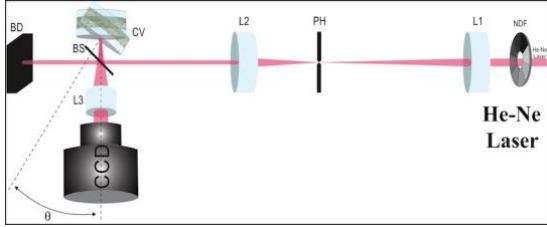

Figure S 4. Experimental setup for determination of the coherent backscattering cone, L1, L2 and L3, lens; PH, pinhole; BS, beam splitter; CV, cuvette of 2 mm optical pathlength; CCD camera; BD, beam dump. The sample (CV) was rotated horizontally 30°, 60° and 70° with respect to the normal incidence, which correspond to incidence angles into the sample of 0° (0 mrad), 19.07° (333 mrad), 34.47° (600 mrad) and 37.89 (661 mrad), respectively. The backscattered intensity was measured as a function of the horizontal collection angle.

From an effective internal reflection for the normal incidence $IR(0°)\approx3\%$, we can estimate the absolute intensity of backscattering cone before being reflected at the interface sample–silica ($I^*_{CBC}$). Consequently, $IR(\theta)$ for incidence angles of 30°, 60° and 70°, can be determined by a simple mathematical relation $IR(\theta)=1-I_{CBC}(\theta)/I^*_{CBC}$, where $I_{CBC}(\theta)$ is the measured intensity of backscattering cone for each incidence angle $\theta$. The effective internal reflection undergone by the coherently backscattered photons, $IR(0°)\approx3\%$, $IR(30°)\approx20\%$, $IR(60°)\approx45\%$ and $IR(70°)\approx65\%$, is considerably higher than the specular reflection measured for the photons that enter the sample (<1%) in the exact opposite direction. This indicates a non–reciprocal propagation of light, i.e. mirror–symmetry (parity symmetry) breaking. Notice that this large increase in the effective internal reflection undergone by the photons leaving the sample (coherently backscattered) would only be possible if the effective refractive index felt by these photons is largely enhanced. For a classical refractive index (1.53), the internal reflection for the photons leaving the sample (sample–silica interface) would be <1% for both polarizations and all incidence angles. Notice that for incidence angles $\theta$ of 0°, 30°, 60° and 70°, the effective incidence angles for the coherently backscattered photons at the interface sample–silica correspond to 0°, 19.07°, 34.47° and 37.89°, respectively.

The transport mean free path was extracted from the half angle of the backscattering cone (figure 3g, right) [50]. A simple model for internal reflection was taken into account for correction of the transport mean free path, considering the effective refractive index for depth near zero ($l_{T0}$) [51,52]. In order to estimate the internal reflection, different effective refractive indexes were estimated for each incidence angle. The effective refractive index for each incidence angle for depth near zero ($n_{eff0}(\theta)$) were calculated by the equation S6.

$$n_{eff0}(\theta) = 1 + (n_{eff} - 1)\gamma_0(\theta) \quad \text{S6}$$

Where $\gamma_0(\theta)$ would be the enhanced absorption factor for each incidence angle $\theta$ for depth near zero. $n_{eff}$ is equal to 1.53. $\gamma_0(\theta)$ are estimated by the equation S7.

$$\gamma_0(\theta) = \gamma_0 \times G(\infty;\theta) \quad \text{S7}$$

Where $G(\infty;\theta)$ are the asymptotic values of relative conductance, extracted by the transmission and propagation experiments (figure 1c and 2a), for each incidence angle $\theta$. Notice that for negligible absorption, $G(\infty;\theta)$ values would represent the enhancement factor of localization for depth near zero for each incidence angle $\theta$. Table SI shows the asymptotic value of relative conductance (extracted from the propagation experiment), the enhanced absorption factor for depth near zero (determined by equation S7), the effective refractive index (determined by equation S6) and $l_{T0}$ extracted from the half angle of backscattering cone and corrected by the internal reflection, for incidence angles of 0°, 30°, 60° and 70°.

**Table S I.** For each incidence angle $\theta$ (0°, 30°, 60° and 70°): asymptotic value of relative conductance ($G(\infty;\theta)$), The enhanced absorption factor ($\gamma_0(\theta)$) and the effective refractive index ($n_{eff0}(\theta)$) for depth near zero, and the transport mean free path corrected by internal reflection considering the effective refractive index $n_{eff0}(\theta)$.

| $\theta$ | $G(\infty;\theta)$ | $\gamma_0(\theta)$ | $n_{eff0}(\theta)$ | Half angle CBC (mrad) | $l_{T0}$ (μm) |
|---|---|---|---|---|---|
| 0° | 1 | 1.84 | 1.98 | 54 | 0.86 |
| 30° | 1.06 | 1.95 | 2.03 | 67 | 0.71 |
| 60° | 1.17 | 2.15 | 2.14 | 96 | 0.48 |
| 70° | 1.28 | 2.35 | 2.25 | 140 | 0.32 |

As can be observed in figure 3f, $l_{T0}$ tends to be zero when the effective internal reflection $IR(\%)\rightarrow100\%$, which would be congruent. We must highlight that the correction of internal reflection has been carried out according to the formalisms described by et. al [51] and Zhu et. al [52] for diffusive regime. However, at localization a more complex phenomenon could take place, which would require a new theoretical approach for $l_{T0}$ correction by the internal reflection. Additionally, the $n_{eff0}(\theta)$ correction and, consequently, $l_{T0}$ values, could be potentially affected by inaccurate values of relative conductance, which in turn would be affected by absorption near the input border.

## 8. REFERENCES

1. S. John, "Electromagnetic absorption in a disordered medium near a photon mobility edge," Phys. Rev. Lett. **53**(22), 2169–2172 (1984).
2. P. W. Anderson, "The question of classical localization A theory of white paint?," Philos. Mag. Part B **52**(3), 505–509 (1985).
3. S. John, "Strong localization of photons in certain disordered dielectric superlattices," Phys. Rev. Lett. **58**(23),

4. S. John, "Localization of Light," Phys. Today **44**(5), 32 (1991).
5. M. A. Noginov, G. Zhu, A. M. Belgrave, R. Bakker, V. M. Shalaev, E. E. Narimanov, S. Stout, E. Herz, T. Suteewong, and U. Wiesner, "Demonstration of a spaser-based nanolaser," Nature **460**(7259), 1110–1112 (2009).
6. E. Jimenez-Villar, I. F. da Silva, V. Mestre, P. C. de Oliveira, W. M. Faustino, and G. F. de Sá, "Anderson localization of light in a colloidal suspension (TiO 2 @silica)," Nanoscale **8**(21), 10938–10946 (2016).
7. E. Jimenez Villar, M. C. S. Xavier, J. G. G. S. Ramos, N. U. Wetter, V. Mestre, W. S. Martins, G. F. Basso, V. A. Ermakov, F. C. Marques, and G. F. de Sá, "Localization of light: beginning of a new optics," in *Complex Light and Optical Forces XII*, D. L. Andrews, E. J. Galvez, and J. Glückstad, eds. (Proceeding SPIE 10549, 2018), p. 1054905.
8. R. A. R. Ioffe, A.F., "Non-crystalline, amorphous and liquid electronic semiconductors," Progr. Semicond. **4**, 237–291 (1960).
9. L. Bressel, R. Hass, and O. Reich, "Particle sizing in highly turbid dispersions by Photon Density Wave spectroscopy," J. Quant. Spectrosc. Radiat. Transf. **126**, 122–129 (2013).
10. J.-P. Bouchaud and A. Georges, "Anomalous diffusion in disordered media: Statistical mechanisms, models and physical applications," Phys. Rep. **195**(4–5), 127–293 (1990).
11. P. Tierno, F. Sagués, T. H. Johansen, and I. M. Sokolov, "Antipersistent Random Walk in a Two State Flashing Magnetic Potential," Phys. Rev. Lett. **109**(7), 070601 (2012).
12. E. Abrahams, P. W. Anderson, D. C. Licciardello, and T. V. Ramakrishnan, "Scaling theory of localization: Absence of quantum diffusion in two dimensions," Phys. Rev. Lett. **42**(10), 673–676 (1979).
13. F. Evers and A. D. Mirlin, "Anderson transitions," Rev. Mod. Phys. **80**(4), 1355–1417 (2008).
14. S. E. Skipetrov and J. H. Page, "Red light for Anderson localization," New J. Phys. **18**, 021001 (2016).
15. J. M. Escalante and S. E. Skipetrov, "Longitudinal Optical Fields in Light Scattering from Dielectric Spheres and Anderson Localization of Light," Ann. Phys. **529**(8), 1700039 (2017).
16. D. S. Wiersma, P. Bartolini, A. Lagendijk, and R. Righini, "Localization of light in a disordered medium," Nature **390**(6661), 671–673 (1997).
17. M. Störzer, P. Gross, C. M. Aegerter, and G. Maret, "Observation of the critical regime near anderson localization of light," Phys. Rev. Lett. **96**(6), 1–4 (2006).
18. T. Sperling, W. Bührer, C. M. Aegerter, and G. Maret, "Direct determination of the transition to localization of light in three dimensions," Nat. Photonics **7**(1), 48–52 (2013).
19. F. Scheffold, R. Lenke, R. Tweer, and G. Maret, "Localization or classical diffusion of light?," Nature **398**(6724), 206–207 (1999).
20. F. Scheffold and D. Wiersma, "Inelastic scattering puts in question recent claims of Anderson localization of light," Nat. Photonics **7**(12), 934 (2013).
21. T. Van Der Beek, P. Barthelemy, P. M. Johnson, D. S. Wiersma, and A. Lagendijk, "Light transport through disordered layers of dense gallium arsenide submicron particles," Phys. Rev. B - Condens. Matter Mater. Phys. **85**(11), 1–11 (2012).
22. T. Sperling, L. Schertel, M. Ackermann, G. J. Aubry, C. M. Aegerter, and G. Maret, "Can 3D light localization be reached in "white paint"?," New J. Phys. **18**(1), 13039 (2016).
23. A. A. Chabanov, M. Stoytchev, and A. Z. Genack, "Statistical signatures of photon localization," Nature **404**(6780), 850–853 (2000).
24. E. Jimenez-Villar, V. Mestre, W. S. Martins, G. F. Basso, I. F. da Silva, and G. F. de Sá, "Core-shell TiO 2 @Silica nanoparticles for light confinement," Mater. Today Proc. **4**(11), 11570–11579 (2017).
25. A. F. Demirörs, A. van Blaaderen, and A. Imhof, "Synthesis of Eccentric Titania−Silica Core−Shell and Composite Particles," Chem. Mater. **21**(6), 979–984 (2009).
26. K. Abderrafi, E. Jiménez, T. Ben, S. I. Molina, R. Ibáñez, V. Chirvony, and J. P. Martínez-Pastor, "Production of Nanometer-Size GaAs Nanocristals by Nanosecond Laser Ablation in Liquid," J. Nanosci. Nanotechnol. **12**(8), 6774–6778 (2012).
27. E. Jimenez-Villar, V. Mestre, P. C. de Oliveira, and G. F. de Sá, "Novel core–

shell (TiO2@Silica) nanoparticles for scattering medium in a random laser: higher efficiency, lower laser threshold and lower photodegradation," Nanoscale **5**(24), 12512 (2013).
28. S. E. Skipetrov and I. M. Sokolov, "Absence of Anderson Localization of Light in a Random Ensemble of Point Scatterers," Phys. Rev. Lett. **112**(2), 023905 (2014).
29. E. Jimenez-Villar, V. Mestre, P. C. de Oliveira, W. M. Faustino, D. S. Silva, and G. F. de Sá, "TiO 2 @Silica nanoparticles in a random laser: Strong relationship of silica shell thickness on scattering medium properties and random laser performance," Appl. Phys. Lett. **104**(8), 081909 (2014).
30. E. Rodriguez, E. Jimenez, G. J. Jacob, A. A. R. Neves, C. L. Cesar, and L. C. Barbosa, "Fabrication and characterization of a PbTe quantum dots multilayer structure," Phys. E Low-dimensional Syst. Nanostructures **26**(1–4), 361–365 (2005).
31. E. Rodriguez, G. Kellermann, A. F. Craievich, E. Jimenez, C. L. César, and L. C. Barbosa, "All-optical switching device for infrared based on PbTe quantum dots," Superlattices Microstruct. **43**(5–6), 626–634 (2008).
32. E. Jiménez, K. Abderrafi, R. Abargues, J. L. Valdés, and J. P. Martínez-Pastor, "Laser-Ablation-Induced Synthesis of SiO 2 -Capped Noble Metal Nanoparticles in a Single Step," Langmuir **26**(10), 7458–7463 (2010).
33. E. Jiménez, K. Abderrafi, J. Martínez-Pastor, R. Abargues, J. Luís Valdés, and R. Ibáñez, "A novel method of nanocrystal fabrication based on laser ablation in liquid environment," Superlattices Microstruct. **43**(5–6), 487–493 (2008).
34. J. R. González-Castillo, E. Rodriguez, E. Jimenez-Villar, D. Rodríguez, I. Salomon-García, G. F. de Sá, T. García-Fernández, D. B. Almeida, C. L. Cesar, R. Johnes, and J. C. Ibarra, "Synthesis of Ag@Silica Nanoparticles by Assisted Laser Ablation," Nanoscale Res. Lett. **10**(1), 399 (2015).
35. J. R. González-Castillo, E. Rodríguez-González, E. Jiménez-Villar, C. L. Cesar, and J. A. Andrade-Arvizu, "Assisted laser ablation: silver/gold nanostructures coated with silica," Appl. Nanosci. **7**(8), 597–605 (2017).
36. G. Fuertes, O. L. Sánchez-Muñoz, E. Pedrueza, K. Abderrafi, J. Salgado, and E. Jiménez, "Switchable Bactericidal Effects from Novel Silica-Coated Silver Nanoparticles Mediated by Light Irradiation," Langmuir **27**(6), 2826–2833 (2011).
37. E. Rodríguez, E. Jimenez, L. A. Padilha, A. A. R. Neves, G. J. Jacob, C. L. César, and L. C. Barbosa, "SiO2/PbTe quantum-dot multilayer production and characterization," Appl. Phys. Lett. **86**(11), 113117 (2005).
38. G. Kellermann, E. Rodriguez, E. Jimenez, C. L. Cesar, L. C. Barbosa, and A. F. Craievich, "Structure of PbTe(SiO 2 )/SiO 2 multilayers deposited on Si(111)," J. Appl. Crystallogr. **43**(3), 385–393 (2010).
39. A. D. Mirlin, "Spatial structure of anomalously localized states in disordered conductors," J. Math. Phys. **38**(4), 1888–1917 (1997).
40. A. Mirlin, "Statistics of energy levels and eigenfunctions in disordered systems," Phys. Rep. **326**(5–6), 259–382 (2000).
41. G. Campagnano and Y. V Nazarov, "G(Q) corrections in the circuit theory of quantum transport," Phys. Rev. B **74**(12), 1–15 (2006).
42. A. L. R. Barbosa, D. Bazeia, and J. G. G. S. Ramos, "Universal Braess paradox in open quantum dots," Phys. Rev. E **90**(4), 042915 (2014).
43. E. Jiménez-Villar, I. F. da Silva, V. Mestre, N. U. Wetter, C. Lopez, P. C. de Oliveira, W. M. Faustino, and G. F. de Sá, "Random Lasing at Localization Transition in a Colloidal Suspension (TiO 2 @Silica)," ACS Omega **2**(6), 2415–2421 (2017).
44. N. U. Wetter, J. M. Giehl, F. Butzbach, D. Anacleto, and E. Jiménez-Villar, "Polydispersed Powders (Nd 3+ :YVO 4 ) for Ultra Efficient Random Lasers," Part. Part. Syst. Charact. **35**(4), 1700335 (2018).
45. M. B. van der Mark, M. P. van Albada, and A. Lagendijk, "Light scattering in strongly scattering media: Multiple scattering and weak localization," Phys. Rev. B **37**(7), 3575–3592 (1988).
46. B. L. Al'tshuler, I. K. Zharekeshev, S. a Kotochigova, and V. I. Shklovskiĭ, "Repulsion between energy levels and the metal-insulator transition," Zhurnal Eksp. i Teor. Fiz. **67**(3), 343–355 (1988).
47. B. Maes, P. Bienstman, and R. Baets, "Switching in coupled nonlinear photonic-crystal resonators," J. Opt. Soc. Am. B

**22**(8), 1778 (2005).
48. B. Maes, M. Soljacic, J. D. Joannopoulos, P. Bienstman, R. Baets, S.-P. Gorza, and M. Haelterman, "Switching through symmetry breaking in coupled nonlinear micro-cavities," Opt. Express **14**(22), 10678 (2006).
49. P. Hamel, S. Haddadi, F. Raineri, P. Monnier, G. Beaudoin, I. Sagnes, A. Levenson, and A. M. Yacomotti, "Spontaneous mirror-symmetry breaking in coupled photonic-crystal nanolasers," Nat. Photonics **9**(5), 311–315 (2015).
50. E. Akkermans, P. E. Wolf, and R. Maynard, "Coherent Backscattering of Light by Disordered Media: Analysis of the Peak Line Shape," Phys. Rev. Lett. **56**(14), 1471–1474 (1986).
51. A. Lagendijk, R. Vreeker, and P. De Vries, "Influence of internal reflection on diffusive transport in strongly scattering media," Phys. Lett. A **136**(1–2), 81–88 (1989).
52. J. X. Zhu, D. J. Pine, and D. A. Weitz, "Internal reflection of diffusive light in random media," Phys. Rev. A **44**(6), 3948–3959 (1991).
53. O. L. Sánchez-Muñoz, J. Salgado, J. Martínez-Pastor, and E. Jiménez-Villar, "Synthesis and Physical Stability of Novel Au-Ag@SiO2 Alloy Nanoparticles," Nanosci. Nanotechnol. **2**(1), 1–7 (2012).
54. E. Jimenez-Villar, V. Mestre, N. U. Wetter, and G. F de Sá, "Core-shell (TiO2@Silica) nanoparticles for random lasers," in *Complex Light and Optical Forces XII*, E. J. Galvez, D. L. Andrews, and J. Glückstad, eds. (2018), **10549**, p. 105490D–10549–10.
55. M. Büttiker and M. Moskalets, "FROM ANDERSON LOCALIZATION TO MESOSCOPIC PHYSICS," Int. J. Mod. Phys. B **24**(12n13), 1555–1576 (2010).
56. S. E. Skipetrov and B. A. Van Tiggelen, "Dynamics of anderson localization in open 3D media," Phys. Rev. Lett. **96**(4), 2–5 (2006).